\documentclass[%
reprint,
superscriptaddress,
groupedaddress,
showpacs,
nofootinbib,
nobibnotes,
amsmath,amssymb,
prb
]{revtex4-1}

\usepackage{graphicx}
\usepackage{dcolumn}
\usepackage{bm}
\usepackage{color}
\usepackage{times}

\usepackage{ulem}
\newcommand{\be}{\begin{equation}}
\newcommand{\ee}{\end{equation}}

\begin{document}

\preprint{APS/123-QED}

\title{Point defect absorption by grain boundaries in $\alpha$-iron by atomic density function modeling}

\author{O. Kapikranian}
\altaffiliation[Now at ]{ENSICAEN, 6 Bd Mar\'echal Juin, 14050 Caen, France} 
\altaffiliation[Also at ]{Institute for condensed matter physics of the National Academy of Sciences of Ukraine, 79011 Lviv, Ukraine}
\email[Electronic address: ]{oleksandr.kapikranyan@ensicaen.fr}
\author{H. Zapolsky}
\author{R. Patte}
\author{C. Pareige}
\author{B. Radiguet}
\author{P. Pareige}
\affiliation{Groupe de Physique des Mat\'eriaux, Universit\'e et INSA de Rouen, UMR CNRS 6634,
Av. de l'universit\'e, BP 12, 76801 Saint Etienne du Rouvray, France}

\date{\today}

\begin{abstract}
Using the atomic density function theory (ADFT), we examine the point defect absorption at [110] symmetrical tilt grain boundaries in body-centered cubic iron. It is found that the sink strength strongly depends on misorientation angle. We also show that the ADFT is able to reproduce reasonably well the elastic properties and the point defect formation volume in $\alpha$-iron.
\end{abstract}

\pacs{61.50.Ah, 61.72--y, 61.82.Bg, 62.20.D--}
\maketitle


\section{\label{intro} Introduction}

Point defect sinks, such as individual dislocations or grain boundaries, 
play a crucial role in embrittlement, swelling, or non-equilibrium solute segregation 
driven by the point defect fluxes \cite{GaryWas}.
These phenomena are especially important in irradiated materials. Fine grain 
polycrystalline materials can exhibit enhanced resistance to irradiation as they possess high concentration of point defect sinks in the form of grain interfaces \cite{Demkowicz2011}. A few experimental studies have been reported about irradiation effects on nanocrystalline materials, 
which confirm good resistance against irradiation \cite{Radiguet2008,Etienne2011,Sun2014,Alsabbagh2013,Chimi2001}. 

Usually, theoretical determination of GB sink strength is based on the dislocation representation of grain boundaries and, consequently, on the consideration of elastic interaction between the latter and point self defects \cite{KingSmith1981,JiangEtAl2014}. This approach is hardly applicable to an arbitrary GB geometry that can be found in real materials. Molecular statics results for the vacancy and self-interstitial atom formation energies at GBs have appeared in literature recently \cite{Tschopp2012}. Atomistic dynamics simulations remain numerically too costly, and thus limited to individual nanograins \cite{ZhangHuang2012}, since one is dealing with a diffusional phenomena. The atomic density function theory (ADFT), which is not atomistic in the strict meaning of the term, but keeps track of the atomic microstructure, seems to be a good candidate for this task as it operates on diffusional time scale. 
It was shown in Ref.~[\onlinecite{KapikranianEtAl2014}] that the ADFT gives correct atomic configurations for various GB geometries. Basic features of grain boundaries as well as dislocation emission from grain boundaries have also been examined in literature using the phase-field-crystal model, which is very close in spirit to the ADFT. There is, however, a question how to incorporate self point defects in this type of model.

In this paper, a way of describing self point defects in the earlier introduced atomic density function model \cite{KapikranianEtAl2014} is proposed. Using this new development of the ADFT, both vacancy and self-interstitial absorption by grain boundaries is modeled. It is shown that simulation results give access to the sink strengths of GBs of different misorientations. Thus, the approach proposed in the present paper opens a possibility of a unified modeling of polycrystalline materials sink strength.

\section{\label{I} Equilibrium state and elastic properties}

To model the point defect absorption at grain boundaries the ADFT has been used. A theoretical foundation of the ADFT is based on the nonequilibrium Helmholtz free energy of a system that is a functional $F[\rho]$ of an atomic density function, $\rho({\bf r})$. This function is an occupation probability to find an atom at the site ${\bf r}$ of the underlying Ising lattice.
Following Refs.~[\onlinecite{Jin-Khachaturyan-2006}] and [\onlinecite{KapikranianEtAl2014}], the free energy functional can be written as:
\begin{eqnarray}\label{F_tot}
&& F[\rho] = \frac{1}{2}\int \frac{d^3r}{V} \int \frac{d^3r'}{V} W(|{\bf r-r}'|)\rho({\bf r})\rho({\bf r}')
\\\nonumber
&& + k_BT\int \frac{d^3r}{V}\left[\rho({\bf r})\ln\rho({\bf r})+(1-\rho({\bf r}))\ln(1-\rho({\bf r}))\right],
\end{eqnarray}
where $W(|{\bf r-r}'|)$ is the atomic interaction potential and $V$ the system volume.
The form of the Fourier transform $V(k)$ of $W(|{\bf r-r}'|)$ was chosen to reproduce the form of the first peak of the structural factor of iron, calculated by molecular dynamics at the melting point \cite{JaatinenElder2009}:
\be\label{V(k)def}
V(k)=V_0\left(1-k^4/\left((k^2-{k_1}^2)^2+{k_2}^4\right)\right)
\ee
with $k_1\simeq 0.435 k_0$, and $k_2\simeq 0.626 k_0$, where $k_0$ is the minimum position of the potential $V(k)$, and $V_0$ defines the energy scale.

To model the evolution of the atomic density function $\rho({\bf r})$, the kinetic microscopic diffusion equation proposed in Ref.~[\onlinecite{Jin-Khachaturyan-2006}] has been used. The general form of such equation, assuming that the relaxation rate is linearly proportional to the transformation driving force, is:

\be\label{eq_de_diff}
\partial\rho({\bf r})/\partial t = -L\nabla^2\delta F/\delta\rho,
\ee
where $L$ is a constant. Equation (\ref{eq_de_diff}) is conservative with respect to the mean atomic density $\rho_0=\int d^3r \rho({\bf r})/V$. Nonconservative kinetics is obtained if the Laplace operator is dropped in Eq. (\ref{eq_de_diff}). 

To model the kinetics of defect absorption at grain boundaries, Eq. (\ref{eq_de_diff}) in a reduced form was numerically solved using the semi-implicit Fourier scheme \cite{ChenShen1998}. In our simulations the next set of reduced variables was used: reduced distance $x^*=x/a$ ($a$ is the second-nearest-neighbor distance in a bcc lattice), reduced time $t^*=tLV_0/a^2$, reduced energy $F^*=F/V_0$, and temperature $T^*=k_BT/V_0$.

To validate our potential, the elastic constants of the bcc iron have been evaluated. For this purpose, the three characteristic deformations were used: (a) uniform compression/expansion, $x,y,z\to (1-\xi)x,(1-\xi)y,(1-\xi)z$, (b) equal contraction/expansion along two cube edges, $x,y,z\to (1+\xi)x,(1-\xi)y,z$, and (c) pure shear, $x,y,z\to x+\xi y,y,z$. The elastic constants $C_{44}$, $C'=(C_{11}-C_{12})/2$, and the bulk modulus $B=(C_{11}+2C_{12})/3$ have been obtained numerically from the second derivative of the free energy, Eq.~(\ref{F_tot}), with respect to $\xi$. 

\begin{figure}[h]
\center{\includegraphics[width=0.35\textwidth,angle=-90]{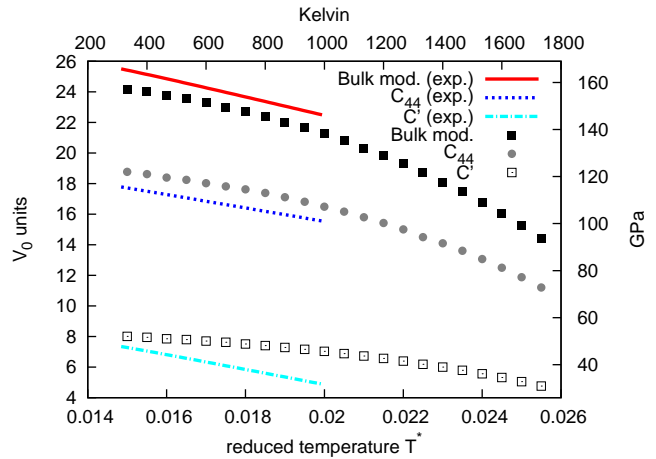}}
\caption{Bcc iron elastic constants in the ADFT as functions of reduced temperature (points). For comparison, the Varshni fits of experimental data from Ref.~[\onlinecite{Adams2006}] are plotted with lines with respect to the upper and right axes.}\label{C44}
\end{figure}

The temperature dependence of the experimental elastic constants of iron is sometimes fitted using the semi-empirical Varshni expression~\cite{Adams2006}. The evolution of the elastic constants of our model with the reduced temperature $T^*$ is presented in Fig. \ref{C44} along with the Varshni fit from Ref.~[\onlinecite{Adams2006}]. The comparison between the experimental and the ADFT data (presented in reduced units) in Fig. \ref{C44} is done by matching two different temperature points in each scale. First, the reduced temperature of solid phase instability in ADFT model is associated with the iron melting temperature (right side of the abscissa axes). Second, we matched the lowest reduced temperature explored in the present study (0.015) and the ambient temperature (300K). This choice, together with the ordinate scaling, has been done to have the best simultaneous fit of the bulk modulus, $C_{44}$ and Zener anisotropy parameter.

Our results follow the general trend of the experimental data, but since our model does not take into account the magnetic ordering, the Zener anisotropy parameter $A=C_{44}/C'$ remains nearly constant through the entire temperature range, contrary to the experiment~\cite{Adams2006}. It is argued, for example in Ref.~[\onlinecite{Razumovskiy2011}], based on DFT calculations, that the temperature dependence of the elastic anisotropy parameter of iron is of magnetic origin. The observed nonlinearity of our results (which leaves the anisotropy parameter $A$ unchanged, though) is due to the vicinity of the stability limit of the solid phases and not due to the magnetic effects.

The Zener anisotropy parameter that we have obtained in our simulations is $A\simeq 2.34$, which is very close to the experimental value of $2.406$ for the bcc iron at room temperature~\cite{Adams2006}. It is important to mention that while an arbitrary GPa scale of the elastic constants can be chosen by assigning a physical value to $V_0$ in the potential Eq. (\ref{V(k)def}), their ratios, and, notably, the anisotropy parameter $A$ are independent from $V_0$ and uniquely determined by the form of the potential derived from the structure factor.

\section{\label{I} Point defects in the ADFT}

To understand how point defects can be introduced in the ADFT, first, the influence of the mean atomic density $\rho_0$ on the form of atomic peaks should be considered. Due to the form of the local free energy in Eq. (\ref{F_tot}) the atomic density function is strictly confined in the interval $[0,1]$. In fact, the local free energy term impedes the growth of inhomogeneities more and more as one approaches 0 or 1, so those values are never actually reached.
During a BCC crystal growth the amplitudes of the peaks related to the atomic positions reach their equilibrium value (close to 1), whereas in between the peaks, the atomic density is approaching zero. This is the reason why with varying the temperature or the density, only the width of the atomic density peaks varies significantly, which is different from the regular phase field crystal (PFC) model. In the PFC description the local term of the free energy is approximated by a Landau polynomial. In this case the amplitude of the atomic peaks is not confined between 0 and 1 but grows with the varying parameters and reaches negative values at its minimum. This difference, as we will show below, is crucial for the elastic properties of the model and vacancy diffusion.

Let us choose as an approximation a Gaussian form of atomic peaks of a fixed height:
\be\label{rho(r)-Gaussian}
\rho({\bf r}) = \sum_i e^{-({\bf r}-{\bf r}_i)^2/\sigma^2},
\ee
with the parameter $\sigma$ controlling the ``width" of the atom; the summation in Eq. (\ref{rho(r)-Gaussian}) is carried out on the sites of a bcc lattice of spacing $a$. It should be noted that the integral of a single Gaussian peak in its Wigner-Seitz primitive cell, normalized to the volume of the cell, gives the mean density $\rho_0$. If the atomic density profile vanishes at the borders of the Wigner-Seitz cell, the integration interval can be extended to infinity and Gaussian integration can be applied, leading to $\sigma=a\left(\rho_0/(2\pi^{3/2})\right)^{1/3}$.

\begin{figure}
\center{\includegraphics[width=0.21\textwidth,angle=-90]{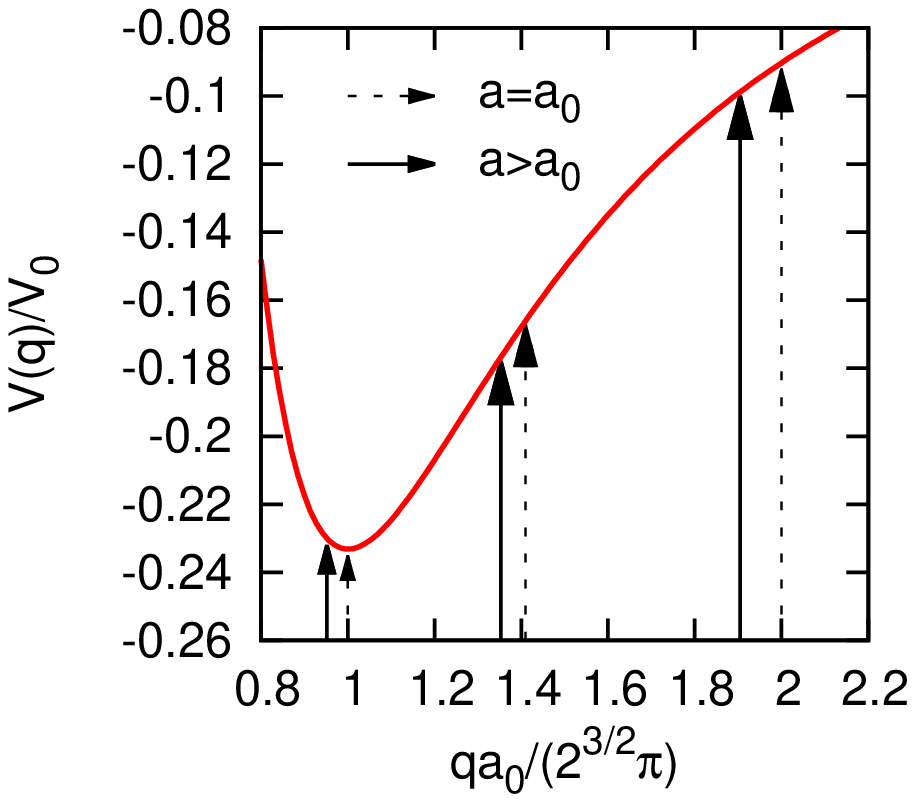}
\includegraphics[width=0.21\textwidth,angle=-90]{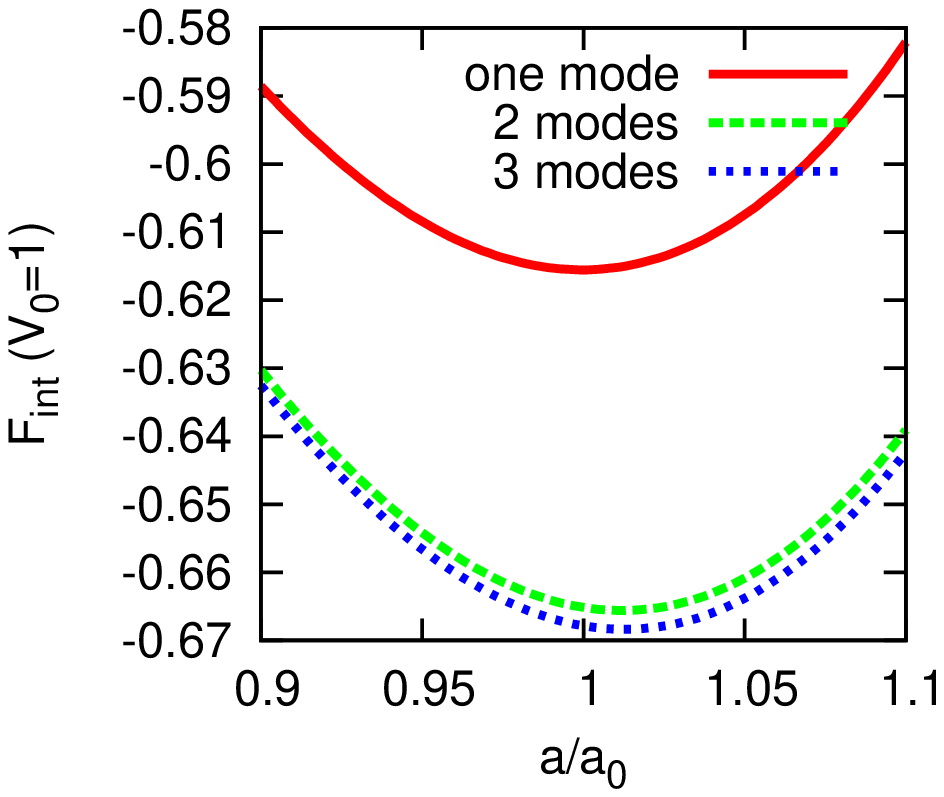}}
\hspace{0.9cm} (a) \hspace{3.7cm} (b)
\caption{(a) The Fourier transform $V(k)$ of the interatomic interaction energy as a function of the distance. (b) Illustration to the equilibrium lattice parameter $a$ following from Eq.(\ref{f_int}) for $\rho_0\simeq 0.0837$. The first three modes positions corresponding to a bcc structure are highlighted with arrows in (a).}\label{modes}
\end{figure}

The Fourier transform of Eq. (\ref{rho(r)-Gaussian}) is
\be\label{rho_k}
\rho_{\bf k} = \rho_0 \delta_{{\bf k},0} + \sum_{{\bf q}}\delta_{{\bf k},{\bf q}}e^{-\sigma^2k^2/4},
\ee
where the sum is over the first Brillouin zone of the reciprocal lattice face-centered cubic lattice. Rewriting Eq. (\ref{F_tot}) in the Fourier variables and using Eq. (\ref{rho_k}), the non-local part, which represents the internal energy $F_\mathrm{int}$, can be written as
\begin{eqnarray}\label{f_int}
F_{\mathrm{int}} &=& \frac{V_0}{2}{\rho_0}^2 + \sum_{{\bf q}}V(q)e^{-\sigma^2{q}^2/2}.
\end{eqnarray}
We will call the $n$-mode the contribution of all the terms in Eq. (\ref{f_int}) with $|{\bf q}|=q_n$, and $q_1<q_2<q_3<\ldots$. The first three modes correspond to $q_1=2\sqrt{2}\pi/a$, $q_2=4\pi/a$, and $q_3=4\sqrt{2}\pi/a$. When the density $\rho_0\ll 1$, which is the case in our simulation, the exponential prefactor in (\ref{f_int}) is nonvanishing for ${\bf q}$-{\it s} corresponding to higher modes. So far, we have made no assumption about the lattice parameter $a$, let us choose $a_0=2^{3/2}\pi/k_0$ [where $k_0$ is the minimum position of the potential $V(k)$] as the reference.
The function $V(k)$ is presented in Fig.~\ref{modes}~(a) where the wave vectors of the first three modes are indicated with arrows (dashed for $a=a_0$ and solid for $a=1.05a_0$). Note that, since $\sigma\sim a$ and $q\sim a^{-1}$, the exponential prefactor in Eq. (\ref{f_int}) does not depend on $a$. The aforementioned figure gives an idea about the energy gain associated with the first mode and its loss associated with higher modes.

The energy given by Eq. (\ref{f_int}) was calculated in one-, two- and three-mode approximations and is plotted in Fig.~\ref{modes}(b) as a function of the ratio $a/a_0$. It can be seen that the minimum of the free energy for two and three modes approximation corresponds to $a_\mathrm{min}>a_0$. Since $\sigma\sim\rho^{1/3}$, the exponential prefactor decreases more slowly with the mode number for lower $\rho_0$, and, consequently, $a_\mathrm{min}$ increases with decreasing $\rho_0$. Then the equilibrium lattice parameter in the ADFT depends on the mean value of the atomic density $\rho_0$ and, consequently, is not uniquely determined by the position $k_0$ of the minimum of the interaction potential $V(k)$. From now on, we will be referring to $a_\mathrm{min}$ as just $a$.

The higher modes contribution is determinant for the elastic properties of the ADFT. If the atomic density profile were not confined between 0 and 1, the first mode would prevail over the following ones and the one mode approximation could have been used. This is equivalent to considering only the lowest wave vectors in Eq. (\ref{f_int}) and putting $\sigma=0$ (the first mode amplitude being normalized to 1). This would lead to $C_{11}=16\alpha$, $C_{12}=C_{44}=8\alpha$, with $\alpha=V_0\left[(k_1/k_2)^4+(k_1/k_2)^8\right]$, and anisotropy parameter $A=2$, which differ significantly from the iron elastic constants relations \cite{Adams2006}. The fact that the ADFT reproduces significantly better the elastic constants of bcc iron, is due to the particular form of the local free energy term in Eq. (\ref{F_tot}) that makes the higher modes contribution important. 

\begin{figure}
\includegraphics[angle=-90,width=0.245\textwidth]{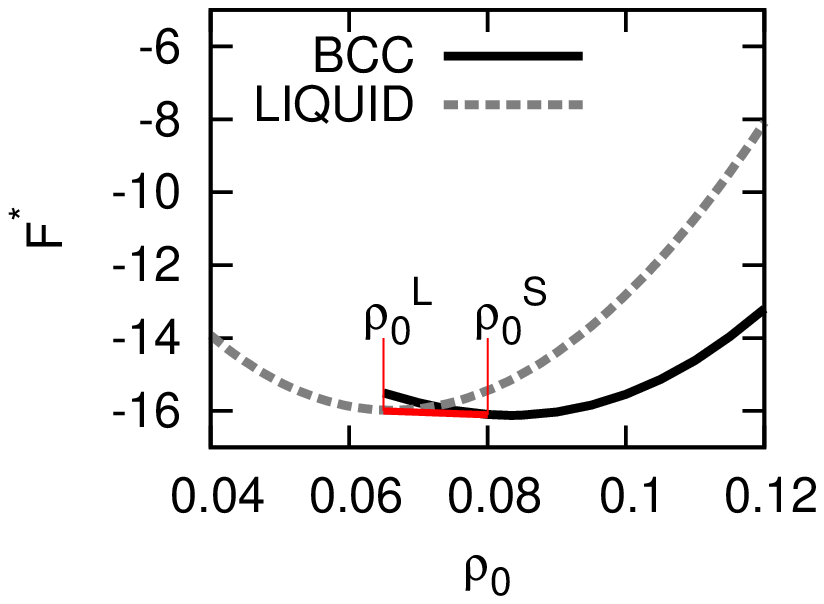}
\includegraphics[angle=-90,width=0.23\textwidth]{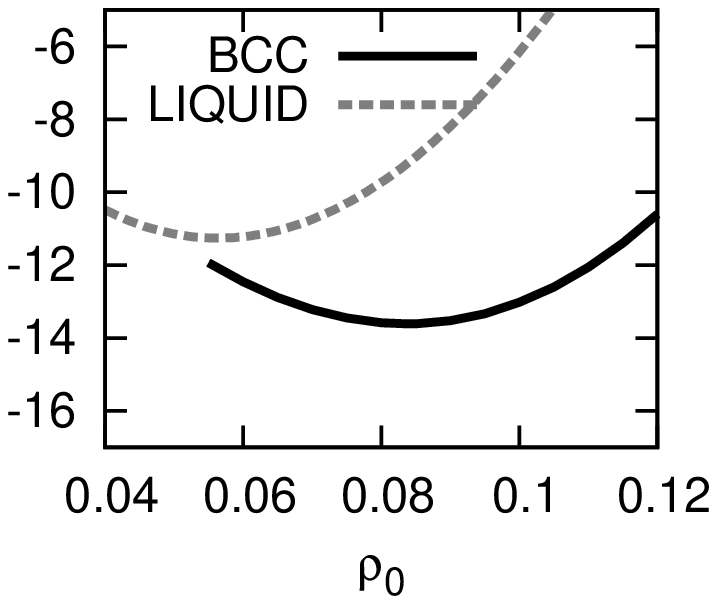}
\center{(a)\hspace{3cm}(b)}
\caption{The solid and liquid state free energies for reduced temperatures $T^*=0.025$ (a) and $0.02$ (b).}\label{common_tang_fig}
\end{figure}

In Fig. \ref{common_tang_fig}, the free energy as a function of $\rho_0$ is presented for two different temperatures. The free energy of the liquid state is given by $F_\mathrm{liq.}=V_0{\rho_0}^2/2+k_BT(\rho_0\ln\rho_0+(1-\rho_0)\ln(1-\rho_0))$, whereas that of the solid phase is computed numerically. The fit of the structure factor used in this paper corresponds to the temperature $T^*\simeq 0.025$ [Fig. \ref{common_tang_fig}(a)]. The equilibrium values of the atomic density in solid and liquid phases were found using a common tangent construction. Assuming that at this temperature the atomic volume of liquid at coexistence with solid is the same as the atomic volume of the metastable solid at the same ${\rho_0}$, one can estimate the melting volume change.

\begin{figure}[h]
\center{\includegraphics[width=0.25\textwidth,angle=-90]{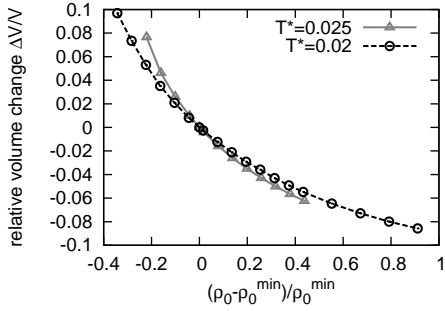}}
\caption{The relative volume change as a function of the average atomic density profile $\rho_0$. The values ${\rho_0}^{\mathrm{min}}$ corresponding to the solid free energy minimum position in Figs.~\ref{common_tang_fig}(a) and \ref{common_tang_fig}(b) are 0.0837 and 0.0834, respectively.}\label{relat_vol_change}
\end{figure}

Thus, instead of the lattice expansion $\Delta a/a=(a(\rho_0)-a(\rho^\mathrm{min}_0))/a(\rho^\mathrm{min}_0)$, we will rather speak of the relative volume change $\Delta V/V=3\Delta a/a$. We have found numerically the dependence of $\Delta V/V$ on $\rho_0$, by relaxing the lattice constant $a$ for each value of the average atomic density (see Fig. \ref{relat_vol_change}). As one would expect, due to thermal expansion, $a({\rho_0}^{\mathrm{min}})$ increases when going from $T^*=0.02$ to 0.025. At the same time, $\rho^\mathrm{min}_0$ slightly decreases (see the legend of Fig. \ref{relat_vol_change}). Unfortunately, this trend does not persist for the whole temperature range, so we cannot make a real comparison to the experimental data on thermal expansion of iron (for example, see Ref.~[\onlinecite{Seki2005}]). From the data corresponding to $T^*=0.025$ in Fig. \ref{relat_vol_change}, and the extreme points of the liquid-solid coexistence region ${\rho_0}^\mathrm{L}$ and ${\rho_0}^\mathrm{S}$ in Fig. \ref{common_tang_fig}, one gets a relative volume change on melting, $\left(\Delta V/V\right)_{\mathrm{melt.}}=3(a({\rho_0}^\mathrm{L})-a({\rho_0}^\mathrm{S}))/a({\rho_0}^\mathrm{S})$ [where $a({\rho_0}^\mathrm{L})$ corresponds to a metastable solid at ${\rho_0}^\mathrm{L}$] of around $6\%$. It is consistent with the general result from Ref.~[\onlinecite{LuJiang2004}] for bcc lattices, following from random packings and the Goldschmidt premise.

Using these results at lower temperatures, when the solid state becomes absolutely stable, the case when $\rho_0<{\rho_0}^{\mathrm{min}}$ can be considered as a crystal with some (non-equilibrium) concentration of vacancies. In the case when $\rho_0>{\rho_0}^{\mathrm{min}}$ the crystal will be said to contain some (non-equilibrium) concentration of self-interstitial atoms (SIA). For example, if one atomic peak removal is associated with a single vacancy, $|\rho_0-{\rho_0}^{\mathrm{min}}|/{\rho_0}^{\mathrm{min}}$ gives the vacancy concentration $c_\mathrm{v}$. The relative volume change due to the vacancies is given by the formula $\Delta V/V=c_\mathrm{v}V^{\mathrm{F}}/\Omega$, where $V^{\mathrm{F}}$ is the vacancy formation volume and $\Omega$ the atomic volume. The linear fit for small $|\rho_0-{\rho_0}^{\mathrm{min}}|/{\rho_0}^{\mathrm{min}}$ (approaching 0 from below) in Fig. \ref{relat_vol_change}, which corresponds to small $c_v$, leads to $V^{\mathrm{F}}\simeq 0.22\Omega$ at $T^*=0.02$. The latter corresponds to a relaxation volume $V^{\mathrm{rel}}=V^{\mathrm{F}}-\Omega \simeq -0.78\Omega$. The data available for comparison are all low or zero temperature results. The most reliable, to our knowledge, experimental relaxation volume for vacancies in iron is of --0.05 at 6 K~\cite{PhysicsOfRadiation}. First-principle calculations at 0 K gave a value of --0.45 in Ref.~[\onlinecite{Korzhavyi1999}] that the authors themselves assumed being overestimated due to the absence of local relaxation in their calculations. However, there exist experimental data suggesting a thermal expansion of point defects that is up to 15 times that of the matrix~\cite{ChhabildasGilder1972,Ganne_vonStebut1979}. Applying a factor of 15 would put the experimental 6K data in good agreement with our estimation. Our result can thus be considered as physically reasonable, but in any case the point defect formation volume will be rather considered as a phenomenological parameter in our model.

\section{\label{II} Point defect absorption by grain boundaries}

According to the previous considerations the introduction of point defects in a perfect crystal with a fixed size of the simulation box will simply increase the energy of the system. However, when point defect sinks are present, such as dislocations or grain boundaries, the system will tend to decrease its energy by pushing the point defects to the sinks.

\begin{figure}[h]
\center{Vacancies \hspace{3cm} SIA}
\center{
\includegraphics[width=0.15\textwidth,angle=0]{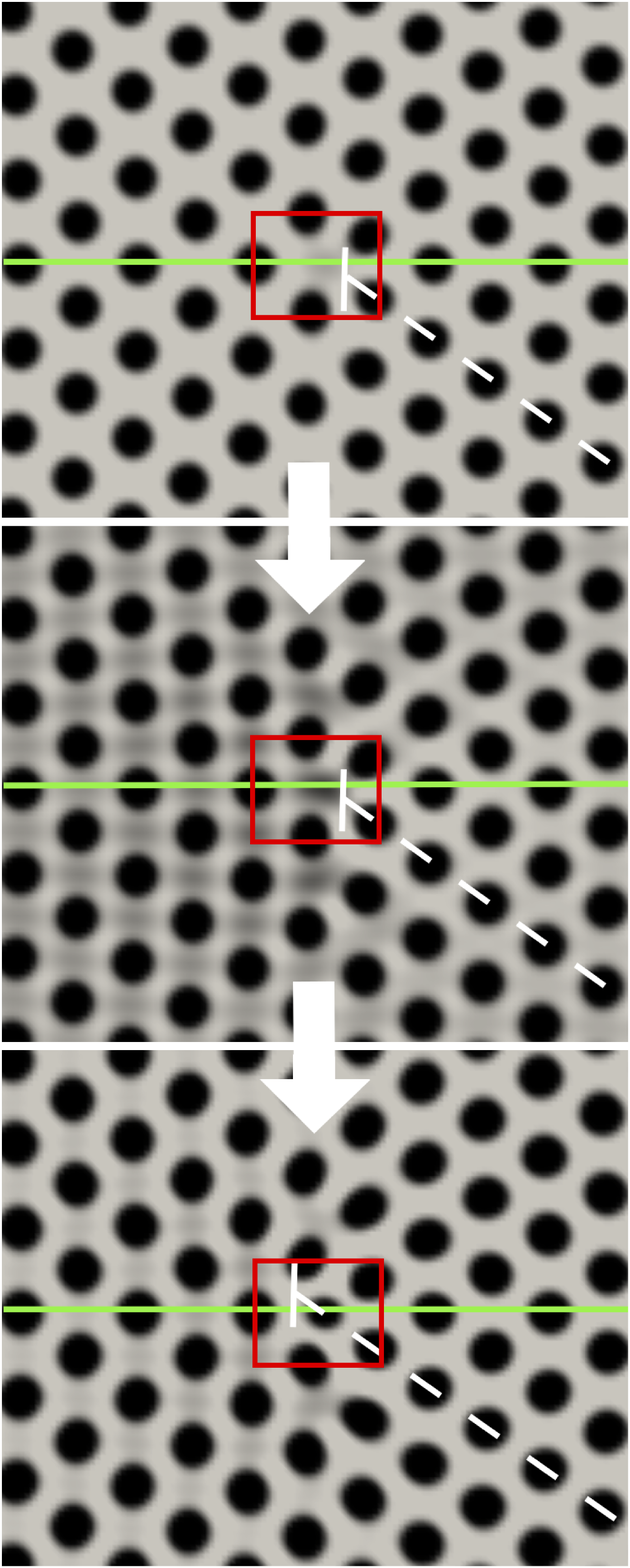}
\includegraphics[width=0.10\textwidth,angle=0]{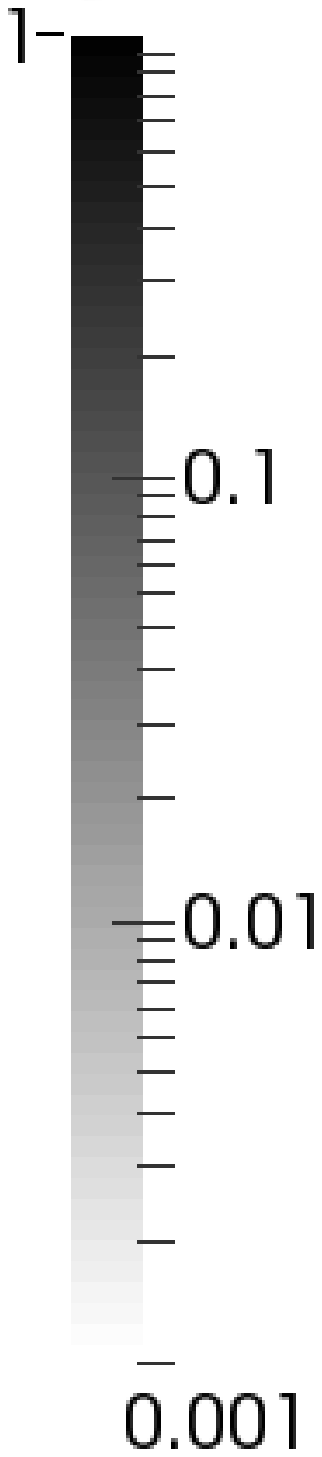}
\includegraphics[width=0.15\textwidth,angle=0]{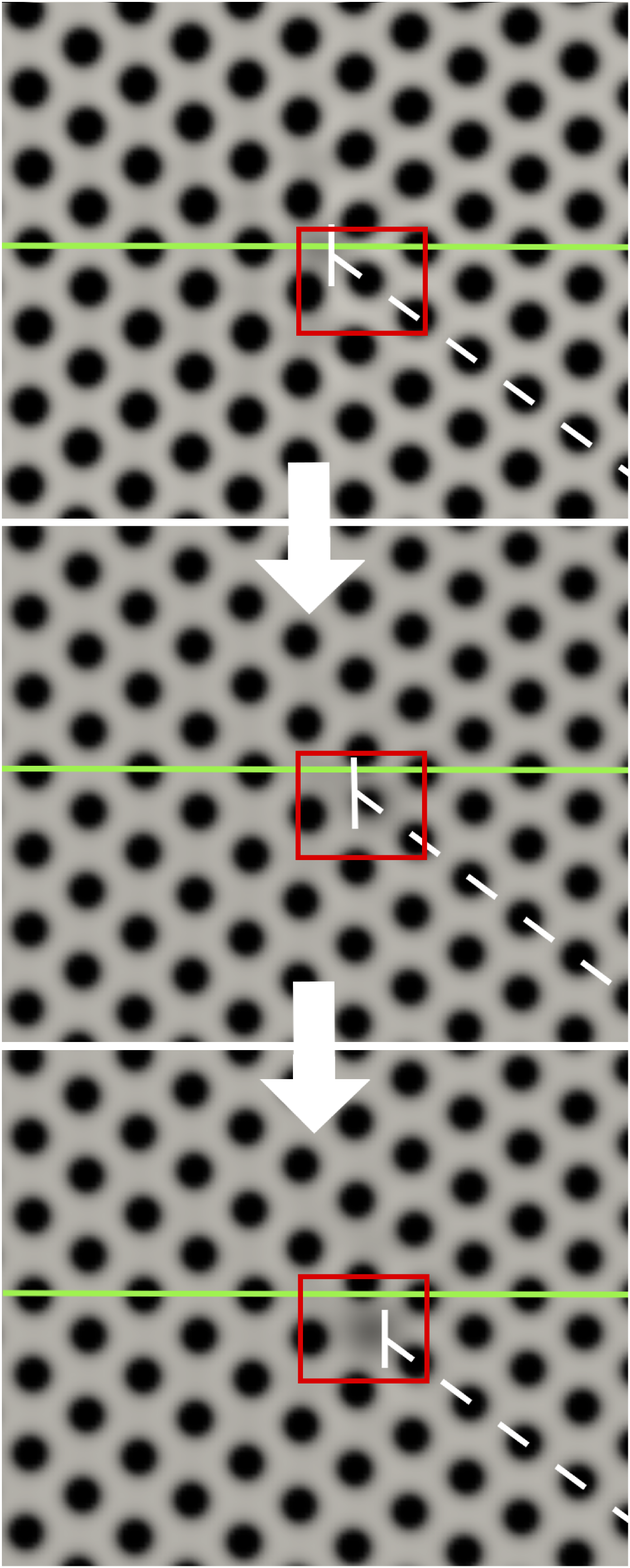}}
\center{(a) \hspace{3cm} (b)}
\caption{Point defect absorption kinetics at a [110] edge dislocation (indicated with white dashed lines) in a system with (a) vacancy and (b) self-interstitial supersaturation. The dislocation is part of a GB which plane is highlighted with green color. The [110] cross-sections of the atomic density function profile are given in logarithmic gray scale in order to make visible the variations of the ADF in between the atomic peaks. The time axis is pointing downwards.}\label{sia_abs}
\end{figure}

To model this phenomena, we have performed simulations with [110] low- and high-angle symmetric tilt grain boundaries by decreasing or increasing the initial equilibrium average density profile value $\rho_0$. The crystal with grain boundaries was constructed using the procedure described in Ref.~[\onlinecite{KapikranianEtAl2014}]. The point defect absorption, manifest in atoms disappearing or appearing at the GB, can be seen directly from Fig. \ref{sia_abs}, where darker colors correspond to higher values of the atomic density function. The relative volume changes used were of $\sim -5\%$ for vacancies and of $\sim 7\%$ for SIA. These values correspond to unrealistically high concentrations of point defects, and compensate for the fact that the grain size is rather small in our simulations ($\sim$ 10--15 nm between neighboring GBs). Indeed, in our simulations both the point defects and the GBs  are by far more numerous than in real materials. This can be alternatively interpreted as artificially increasing the strength of the point defect displacement field. As this increase is the same for all GBs considered, nothing prevents us from determining the relative point defect absorption rates for different types of GBs.

\begin{figure}
\center{
\includegraphics[width=0.05\textwidth,angle=0]{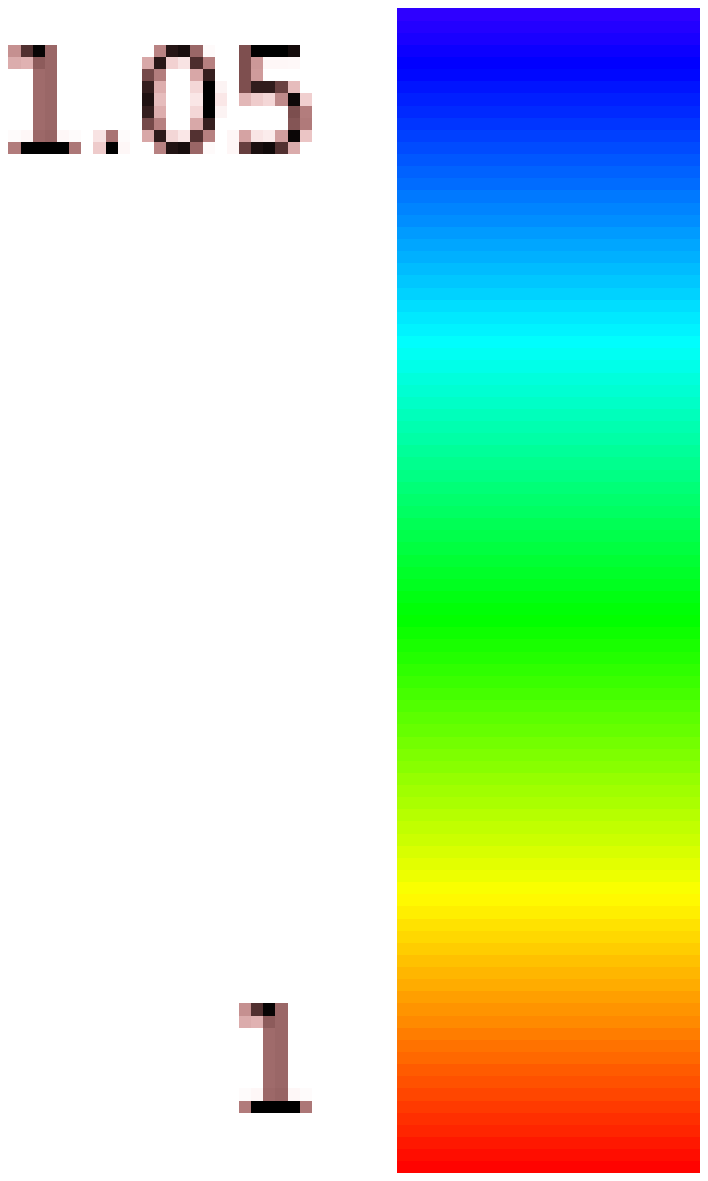}
\includegraphics[width=0.18\textwidth,angle=0]{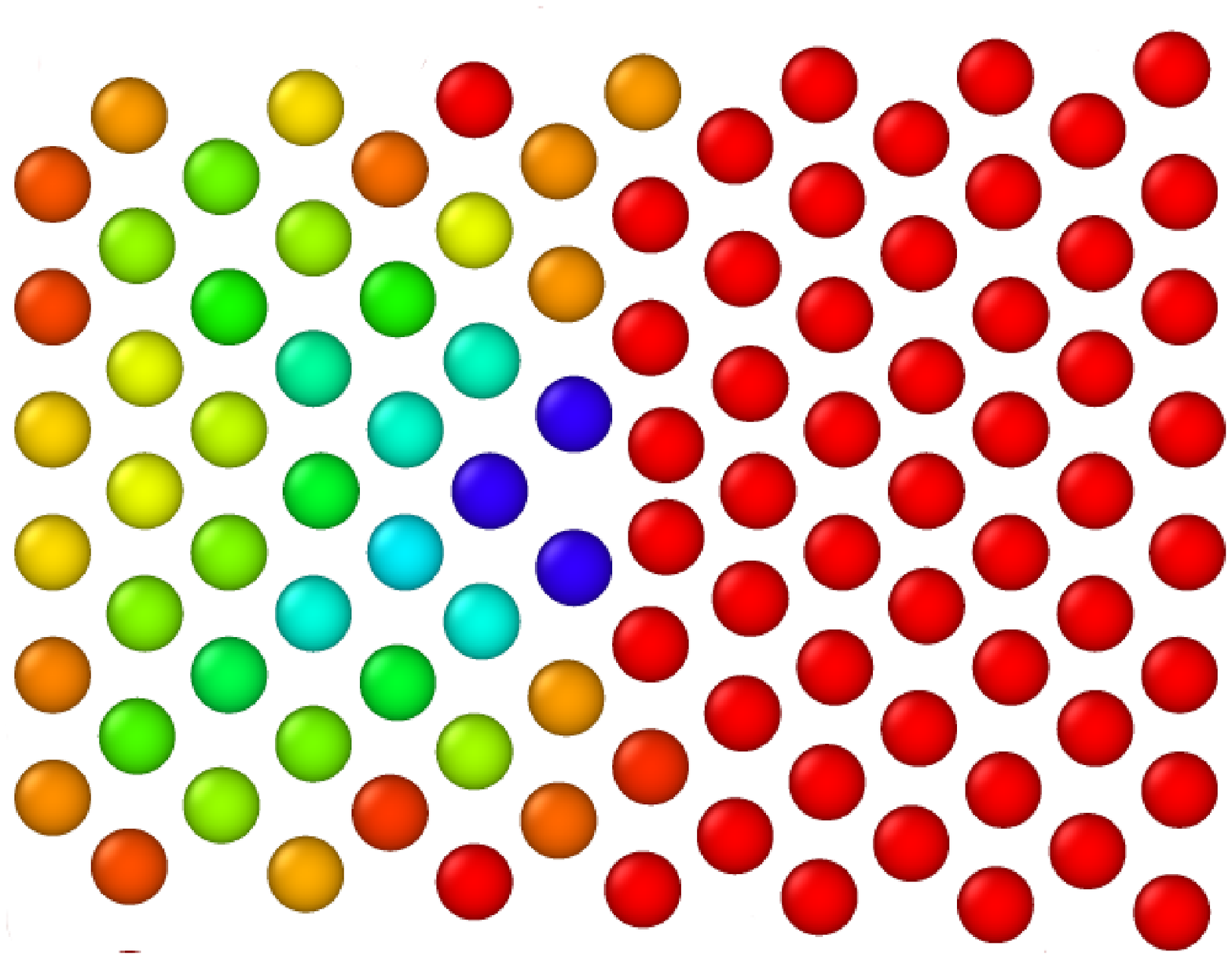}
\includegraphics[width=0.18\textwidth,angle=0]{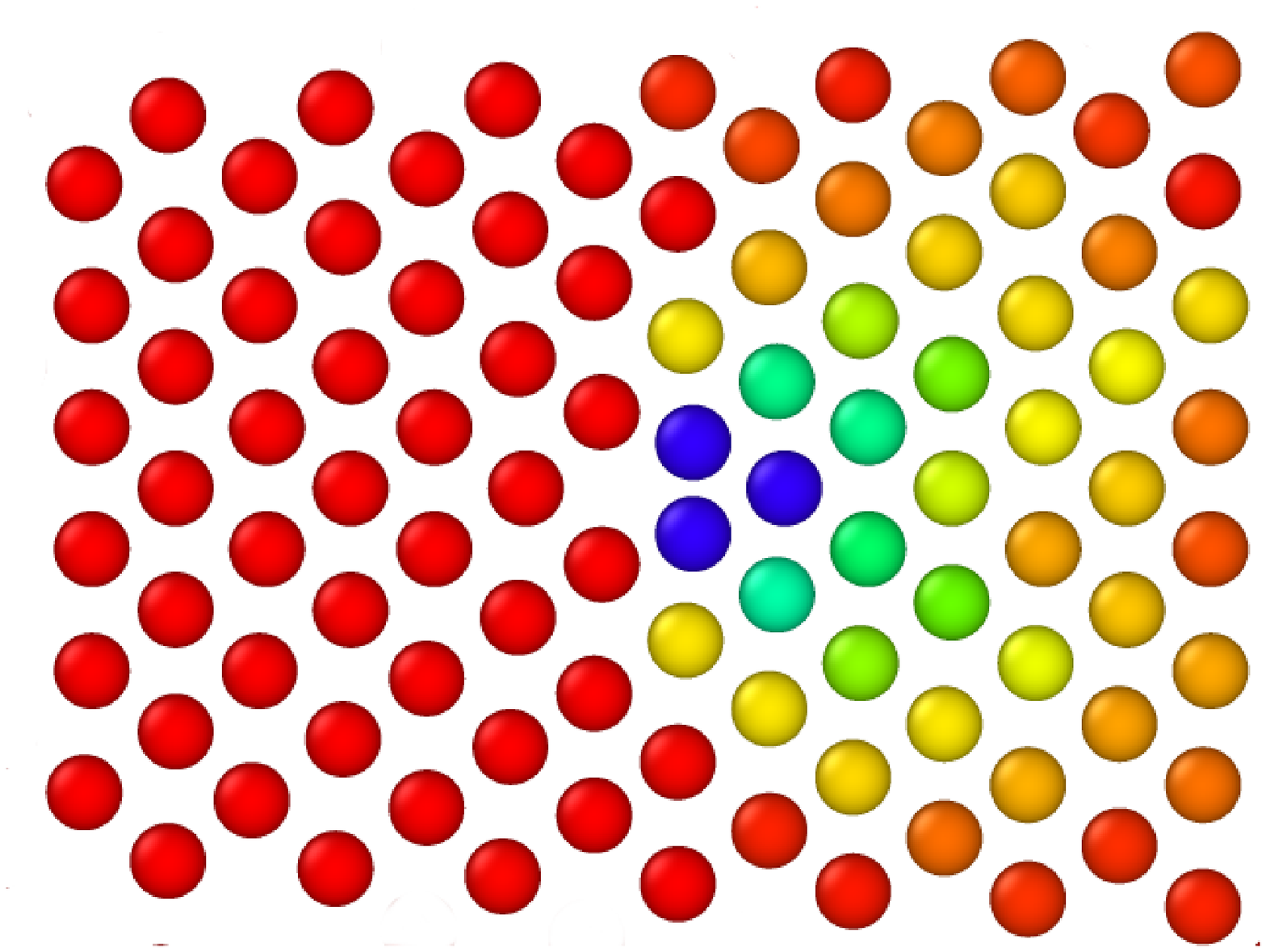}
\includegraphics[width=0.05\textwidth,angle=0]{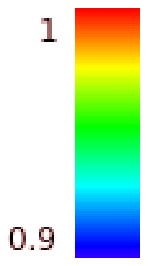}}
\center{(a) \hspace{3cm} (b)}
\caption{The intrinsic strain [dilatation (a) and compression (b)] due to the edge dislocation at the 4.24$^{\circ}$ [110] GB, as given by the local atomic density normalized to its bulk value.}
\label{local_pressure}
\end{figure}

\begin{figure}
(a) \includegraphics[width=0.30\textwidth,angle=-90]{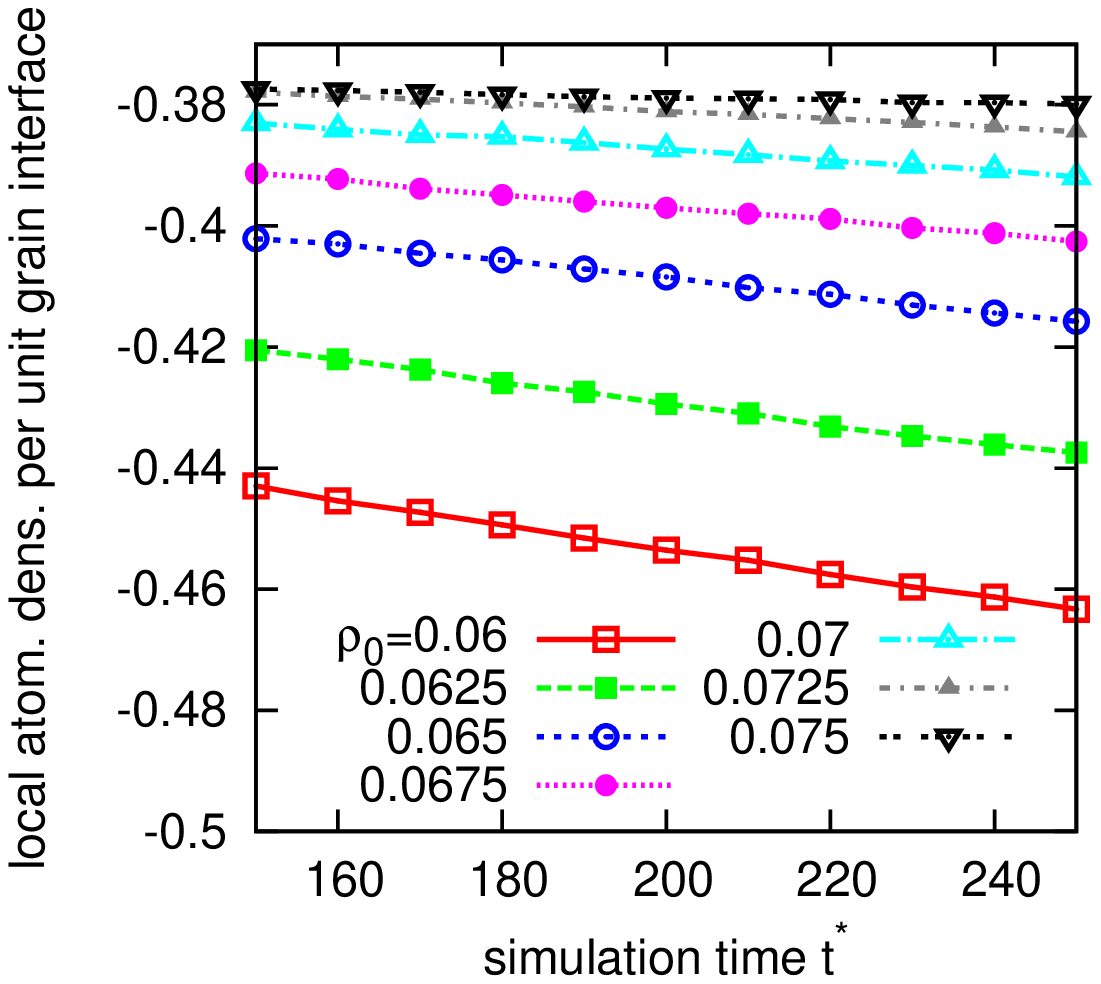}\\
(b) \includegraphics[width=0.27\textwidth,angle=-90]{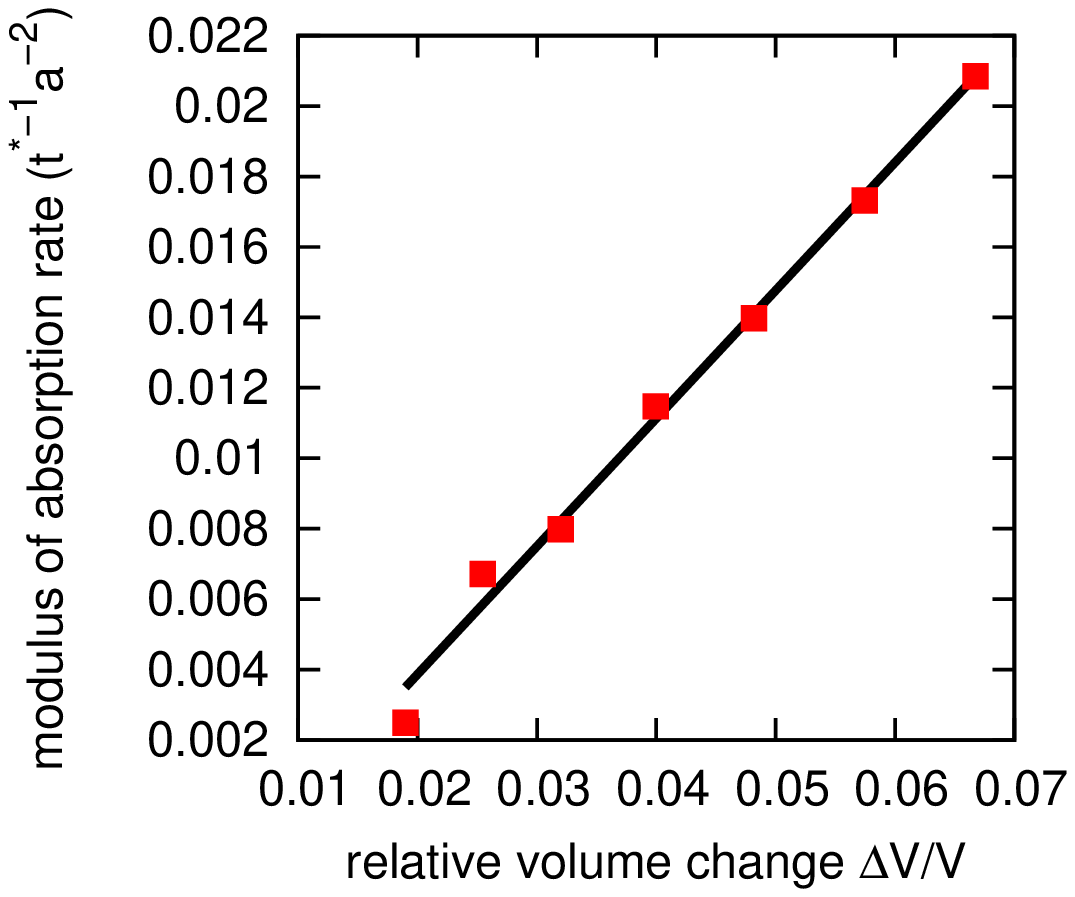}
\caption{The absorption kinetics at the 4.24$^{\circ}$ [110] GB (a), as given by the total decrease of the normalized local atomic density per $a^2$ grain interface. The absolute value of the slope of the curves presented in (a) is plotted in (b) as a function of the relative volume change.}\label{vac_absorp}
\end{figure}

To characterize the volume of an atom, the integration of the ADF in the Voronoi cell associated with this atom has been done. The Voronoi cell is defined as the part of the space situated closer to a given atom than to any other one. This quantity will be referred to as the local atomic density. Then, the relative deviation of the local atomic density from its bulk value will reflect the local strain (compression-expansion) field.
The latter is represented in Fig.~\ref{local_pressure} for a low-angle GB fragment from Fig. \ref{sia_abs}. For better perception of the local atomic density variation, in Fig.~\ref{local_pressure}, colors on equal-sized spheres are used to visualize the deviation of the local atomic density from its bulk value (red color associated to this value is used as the upper rendering threshold for compression and the lower rendering threshold for expansion). The total relative decrease (per unit interface) of the local atomic density in time is plotted in Fig. \ref{vac_absorp}(a) for different tilt angle GBs. The curves are linear, so their slope, plotted in Fig. \ref{vac_absorp}(b) versus the relative volume change, can be taken as the measure of the vacancy absorption rate $\partial c_{\mathrm{v}}/\partial t$. The volume change is itself proportional to the vacancy concentration $c_{\mathrm{v}}$: $\Delta V/V=c_{\mathrm{v}}V^{F}/\Omega$. The fact that the dependence in Fig. \ref{vac_absorp}(b) is nearly linear is consistent with the linear rate equations commonly used to describe point defect absorption by extended defects \cite{DoanMartin2003}.

\begin{figure}
\center{\includegraphics[width=0.3\textwidth,angle=-90]{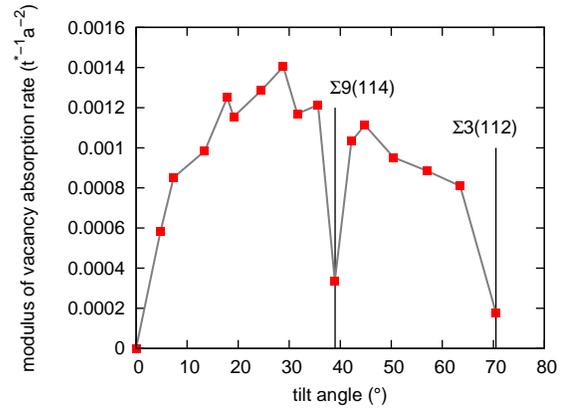}}
\caption{The modulus of the vacancy absorption rate per unit grain interface ($a^2$) as a function of the GB tilt angle.}\label{sink_strength}
\end{figure}

There is no vacancy production {\it during} our simulation, so the rate equation for the vacancy absorption reads as: $\partial c_{\mathrm{v}}/\partial t = - k_{\mathrm{v}}^2D_{\mathrm{v}}c_{\mathrm{v}}$. Since the bulk vacancy diffusion coefficient $D_{\mathrm{v}}$ does not depend on the GB geometry, the vacancy absorption rates of different [110] symmetric tilt grain boundaries plotted in Fig. \ref{sink_strength} reveal their relative sink strengths $k_{\mathrm{v}}^2$. The sink strength of the low-angle GBs increases with the tilt angle due to the increasing of the dislocation density like it was found in Ref.~[\onlinecite{KingSmith1981}]. In Ref.~[\onlinecite{JiangEtAl2014}], however, it was obtained that the sink strength remains almost constant above $3^{\circ}$ tilt, due to the mutually annihilating elastic fields of neighboring dislocations. It was assumed therein that the long-range diffusion of point defects toward the GB is the rate-limiting step for the sink strength determination. 
This assumption does not hold in our model and it is rather the interaction of point defects with the dislocation cores and not with the far-reaching elastic field that is determinant for the absorption rates in our simulations.

It is notable that a low sink strength is obtained for the $\Sigma 9$ and $\Sigma 3$ grain boundaries. This result is in agreement with the experimental study presented in Ref.~[\onlinecite{Watanabe2000}]. The molecular statics modeling done in Ref.~[\onlinecite{Tschopp2012}] is also in favor of our results as it shows a general trend of a mean vacancy formation energy decreasing with $\Sigma$ (that is, the low-$\Sigma$ GBs being the less energetically favorable for absorbing vacancies). Finally, it is coherent with the other properties of special GBs approaching those of the bulk material (energy, excess volume) due to their bulk-like atomic arrangements.

\section{\label{Conclusion} Conclusions}

It has been previously shown that the atomic density function model reproduces well the atomic structures of [100] and [110] symmetric tilt GBs \cite{KapikranianEtAl2014}. 

In the present paper, we have demonstrated that point defects can be also adequately described by this model as well as the point defect absorption by GBs. It comes possible from the adequate description of the elastic properties of given materials. We show that the form of the local term of the free-energy functional plays a crucial role in the description of these properties. 

This new development of the ADFT gives a new insight into the GB sink strength for vacancy annihilation. Accurate fit of the interaction potential to a structural factor allows us to give a quantitative description of vacancy migration to GBs. This gives access to sink strengths of GBs of complex geometries or of polycrystalline materials.

As a perspective, first, Frenkel pairs annihilation in irradiated materials can be modeled with two coupled atomic density functions, one for the vacancies and another one for the self-interstitials. Second, introducing another atomic species, with elastic properties different from the $\alpha$-iron matrix, will allow us to describe non-equilibrium solute segregation at the GBs driven by the fluxes of points defects. This work is in progress.

\section{\label{Acknowledgments} Acknowledgments}

This work was granted access to the HPC resources of IDRIS (under the allocation 2015-097250 made by GENCI) and CRIHAN (under the allocation 2012008). The financial support of Institut Carnot (project MODJI) is acknowledged as well.

\nocite{*}


\begin{thebibliography}{99}

\bibitem{GaryWas} G.S. Was, Fundamentals of Radiation Materials Science: Metals and Alloys, Springer-Verlag Berlin Heidelberg, 2007

\bibitem{Demkowicz2011} M. J. Demkowicz, R. G. Hoagland, B. P. Uberuaga, and A. Misra, Influence of interface sink strength on the reduction of radiation-induced defect concentrations and fluxes in materials with large interface area per unit volume, Phys. Rev. B {\bf 84} (2011) 104102

\bibitem{Radiguet2008} B. Radiguet, A. Etienne, P. Pareige, X. Sauvage, R. Valiev, Irradiation behavior of nanostructured 316 austenitic stainless steel, Journal of Material Science {\bf 43} (2008) 7338-7343

\bibitem{Etienne2011} A. Etienne, B. Radiguet, N.J. Cunningham, G.R. Odette, R. Valiev, and P.Pareige, Comparison of radiation-induced segregation in ultrafine-grained and conventionnal 316 austenitic stainless steels, ULTRAMICROSCOPY {\bf 111} (2011) 659

\bibitem{Sun2014} C. Sun {\it et al.}, Superior radiation-resistant nanoengineered austenitic 304L stainless steel for applications in extreme radiation environments, Sci. Rep. {\bf 5} (2014) 7801

\bibitem{Alsabbagh2013} Ahmad Alsabbagh, Ruslan Z. Valiev, and K.L. Murty, Influence of grain size on radiation effects in a low carbon steel, J. Nucl. Mater. {\bf 443} (2013) 302


\bibitem{Chimi2001} Y. Chimi, A. Iwase, N. Ishikawa, M. Kobiyama, T. Inami, and S. Okuda., Accumulation and recovery of defects in ion-irradiated nanocrystalline gold, J. Nucl. Mater. {\bf 297} (2001) 355

\bibitem{KingSmith1981} A. H. King and D. A. Smith, Calculations of sink strength and bias for point-defect absorption by dislocations in arrays, Radiat. Effects, Volume 54, Issue 3-4, 1981 

\bibitem{JiangEtAl2014} C. Jiang, N. Swaminathan, J. Deng, D. Morgan, and I. Szlufarska, Effect of grain boundary stresses on sink strength, Mater. Res. Lett., 2014, Vol. 2, No. 2, 100–106

\bibitem{Tschopp2012} M. A. Tschopp et al., Probing grain boundary sink strength at the nanoscale: Energetics and length scales of vacancy and interstitial absorption by grain boundaries in $\alpha$-Fe, Phys. Rev. B {\bf 85} (2012) 064108


\bibitem{ZhangHuang2012} Y. Zhang, H. Huang, P.C. Millett, M. Tonks, D. Wolf, and S. R. Phillpot, Atomistic study of grain boundary sink strength under prolonged electron irradiation, J. Nucl. Mater. 422 (2012) 69–76


\bibitem{KapikranianEtAl2014} O. Kapikranian, H. Zapolsky, C. Domain, R. Patte, C. Pareige, B. Radiguet, and P. Pareige, Atomic structure of grain boundaries in iron modeled using
the atomic density function, Phys. Rev. B {\bf 89} (2014) 014111

\bibitem{Jin-Khachaturyan-2006} Y.M. Jin and A.G. Khachaturyan, Atomic density function theory and modeling of microstructure evolution at the atomic scale, J. Appl. Phys. {\bf 100} (2006) 013519

\bibitem{JaatinenElder2009} A. Jaatinen, C. V. Achim, K. R. Elder, and T. Ala-Nissila, Thermodynamics of bcc metals in phase-field-crystal models, Phys. Rev. E {\bf 80} (2009) 031602

\bibitem{ChenShen1998} L. Q. Chen, Jie Shen, Application of semi-implicit Fourier-spectral method to phase field equations, Comput. Phys. Commun. 108 (1998) 147-158

\bibitem{Adams2006} J. J. Adams, D. S. Agosta, R. G. Leisure, and H. Ledbetter, Elastic constants of monocrystal iron from 3 to 500 K, Journal of Applied Physics {\bf 100} (2006) 113530

\bibitem{Razumovskiy2011} V. I. Razumovskiy, A. V. Ruban, and P. A. Korzhavyi, Effect of temperature on the elastic anisotropy of pure Fe and Fe$_{0.9}$Cr$_{0.1}$ random alloy, Phys. Rev. Lett. {\bf 107} (2011) 205504

\bibitem{Seki2005} I. Seki and K. Nagata, Lattice constant of iron and austenite including its supersaturation phase of carbon, ISIJ Int. {\bf 45} (2005) 1789

\bibitem{LuJiang2004} H. M. Lu and Q. Jiang, Melting volume change of different crystalline lattices, Phys. Stat. Sol. (b) {\bf 241} (no. 11) (2004) 2472

\bibitem{PhysicsOfRadiation} P. Ehrhart, K.H. Robrock and H.R. Schober, Basic defects in metals, in: Physics of radiation effects in crystals, Modern Problems in Condensed Matter Sciences, vol. 13 (Elsevier Science Publishers B.V., New York, 1986), p. 63

\bibitem{Korzhavyi1999} P. A. Korzhavyi, I. A. Abrikosov, B. Johansson, A. V. Ruban, and H. L. Skriver, First-principles calculations of the vacancy formation energy in transition and noble metals, Phys. Rev. B {\bf 59} (1999) 11693

\bibitem{ChhabildasGilder1972} L.C. Chhabildas and H.M.Gilder, Thermal coefficient of expansion of an activated vacancy in zinc from high-pressure self-diffusion experiments, Phys. Rev. B {\bf 5} (1972) 2135

\bibitem{Ganne_vonStebut1979} J.P. Ganne and J. von Stebut, Measurement of intrinsic thermal expansion of irradiated defects in aluminum at low temperature, Phys. Rev. Lett. {\bf 43} (1979) 634

\bibitem{DoanMartin2003} N. V. Doan and G. Martin, Elimination of irradiation point defects in crystalline solids: Sink strengths, Phys. Rev. B 67 (2003) 134107

\bibitem{Watanabe2000} S. Watanabe, Y. Takamatsu, N. Sakaguchi, and H. Takahashi, Sink effect of grain boundary on radiation-induced segregation in austenitic stainless steel, J.
Nucl. Mater. {\bf 283-287} (2000) 152


\end{thebibliography}

\end{document}